# Multi-Constraint Molecular Generation using Sparsely Labelled Training Data for Localized High-Concentration Electrolyte Diluent Screening

Jonathan P. Mailoa,[1*] Xin Li,[1] Jiezhong Qiu,[1] and Shengyu Zhang[2*]

1) Tencent Quantum Laboratory, Tencent, Shenzhen, Guangdong, China
2) Tencent Quantum Laboratory, Tencent, Hong Kong SAR, China

[*] corresponding author: jpmailoa@alum.mit.edu, shengyzhang@tencent.com

# Multi-Constraint Molecular Generation using Sparsely Labelled Training Data for Localized High-Concentration Electrolyte Diluent Screening

**Abstract**

Recently, machine learning methods have been used to propose molecules with desired properties, which is especially useful for exploring large chemical spaces efficiently. However, these methods rely on fully labelled training data, and are not practical in situations where molecules with multiple property constraints are required. There is often insufficient training data for all those properties from publicly available databases, especially when ab-initio simulation or experimental property data is also desired for training the conditional molecular generative model. In this work, we show how to modify a semi-supervised variational auto-encoder (SSVAE) model which only works with fully labelled and fully unlabelled molecular property training data into the ConGen model, which also works on training data that have sparsely populated labels. We evaluate ConGen's performance in generating molecules with multiple constraints when trained on a dataset combined from multiple publicly available molecule property databases, and demonstrate an example application of building the virtual chemical space for potential Lithium-ion battery localized high-concentration electrolyte (LHCE) diluents.

## Introduction

Conditional molecular generation capability is a topic of strong interest for the purpose of chemical space exploration in the material virtual screening effort. Efforts in the field of conditional molecular generative model either takes no conditional constraint on the generation approach[1–5] or fail to introduce a cost function based on the generated molecules' property accuracy, making the models' generated molecular properties vary over a large range far from the desired property range.[6] This difficulty arises because in a model, molecular properties are typically the output of some regression model using the molecular structure as input. This makes it more challenging to use molecular properties as the input to conditionally constrain the chemical space of the generated molecules. Recent work based on reinforcement learning has enabled a conditional molecule generator which generates good molecular candidates after thousands of training iterations, assuming that a molecule property evaluator (cheminformatics library or computational material simulation tool) can continuously be utilized on the generated molecules during training.[7] In this work, we are interested in a specific practical task more commonly encountered in the virtual screening of chemical space relevant to industry: given a limited and often incomplete set of molecular property training labels from multiple sources, develop a generative model to generate a molecular chemical space which satisfies multiple property constraints so that it can be used as the high-quality input for a virtual screening pipeline in a low-cost and relatively accurate manner, without requiring additional simulations or experiments to further refine the generative model.

Recent work such as the semi-supervised variational auto encoder (SSVAE) model developed by Kang, *et al*[8,9] solves a part of this problem by employing a dual-track architecture where the molecular property $y$ is simultaneously the output from a molecule regression predictor sub-model and the input to a molecule generation decoder sub-model, in addition to also being the input for a separate molecule encoder sub-model. Because $y$ is an output of the predictor sub-model, it can still be used to construct a useful cost function for the entire model even though it is also being used as the input to control the decoder's generated molecule structures. The resulting combined model has a relatively good control over the generated molecules' property, making it attractive for efficiently generating conditionally constrained molecular chemical space of interest. In addition to that, the SSVAE model is capable of utilizing both fully labelled molecules and fully unlabelled molecules during the training process, making it somewhat attractive for practical usage as there are many cases where we have no access to the molecule properties due to a lack of simulation or experimental data. Nevertheless, the model is still impractical because in practice there are a lot of molecules where the data is only partially labelled and the SSVAE model is not equipped to handle such cases. A practical example of this problem is a situation in battery electrolyte molecule screening where 'easy' molecular properties such as molecular weight (*Mol.Wt*) and the number of fluorine atoms ($n_F$) are easily obtainable from cheminformatics libraries, while simulation data such as ionization energy (*IE*) and experimental data such as the viscosity (*Log.Vis*, or the logarithm of viscosity) are not widely available. If we are interested in generating a chemical space satisfying a number of of these constraints, many of the molecules found in publicly available databases cannot be used as the fully labelled training data for the SSVAE model. Removing the labels completely and turning them into fully unlabelled SSVAE training data is detrimental as we then lose significant valuable label information from our training dataset.

In this work, we show how to enable a generative model which fully utilizes molecules with incomplete labels as the training data for a generative model without having to request additional training data label during training. This model improvement is enabled by modifying the SSVAE model to stop differentiating between fully labelled or unlabelled molecules. The model now relies on a molecular property mask instead, which tells the model which property can be used for training from a given molecule and which cannot. We name this modified SSVAE approach as the ConGen model, and the major modifications needed to enable these practical capabilities will be outlined in the next section. When the supplied molecule training data is either fully labelled or fully unlabelled, the ConGen model's data workflow will look identical to that of the SSVAE model's fully labelled and fully unlabelled data workflow. However, when the ConGen model is supplied with molecules with sparsely populated property labels as the training data, its components and cost functions are appropriately modified such that it only uses the relevant property labels based on the property mask. We first benchmark the usage of this model on a training dataset used by the original SSVAE model, which contains just labelled and unlabelled molecules. We then demonstrate several use cases which cannot be done using the SSVAE model, including the generation of virtual screening chemical space for Lithium-ion battery localized high concentration electrolyte diluent (LHCE) candidates. This is achieved by combining five publicly available molecular property databases, comprising different properties such as *Mol.Wt*, number of fluorine and oxygen atoms ($n_F$ and $n_O$), ionization energy and electron affinity (*IE* and *EA*), and *Log.Vis*. The availability of these properties are very different, with the first three being fully available ('easy'), the next two with availability of approximately 3% ('medium' property, obtainable from quantum chemistry simulations), and the last one with availability of approximately 0.03% ('hard' property, obtainable from experimental measurements).

## Baseline SSVAE Model

We first describe the inner workings of the baseline SSVAE model developed by S. Kang, *et al*,[8] which forms the foundation of this work. The main idea of the SSVAE model is simple:

1. Encode the input molecule structure $x$ from the training dataset into a latent space representation $z$ using an encoder sub-model.

2. Predict the property label of the input molecule structure $x$ from the training dataset into predicted property $y_P$ using a predictor sub-model. If an actual molecule property label $y_L$ exists in the training database, $y_P$ is discarded and the model uses the internal molecule property label $y = y_L$. Otherwise, $y = y_P$ is used.

3. Use the internal molecule property label $y$ and the latent space representation $z$ as input to the decoder sub-model to generate the output molecule structure $x_D$.

In order to handle different types of training data (labeled vs unlabeled), the SSVAE model treats the two types of data differently. The training dataset in an epoch's minibatch is split into two different minibatch (labeled vs unlabeled). The SSVAE workflow is then run twice, in a slightly different manner depending on whether the molecule minibatch is fully labeled or fully unlabeled (**Figure 1**).

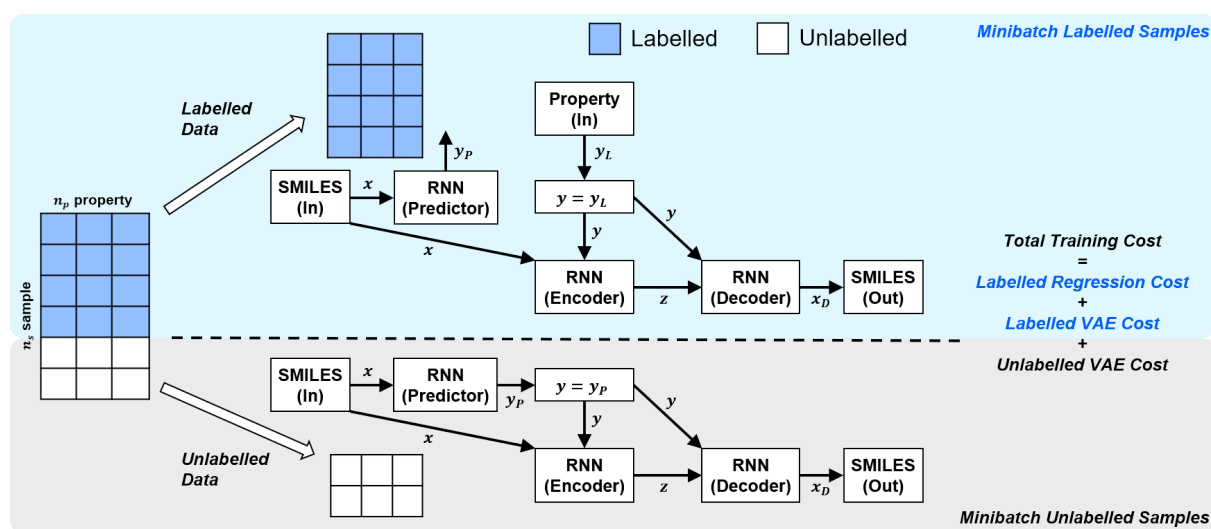

**Figure 1 | High-level labelled / unlabelled data & model differentiation within Kang *et al*'s original SSVAE model.**[8] The variational auto-encoder (VAE) cost is calculated separately for the unlabeled and the labeled dataset, while regression cost is only calculated for the labeled dataset. The three costs are then summed up to calculate the total minibatch training cost.

In SSVAE approach, a molecule entry's training cost function needs to be split into three parts (**Equation 1-3**). The cost function is written in verbose detail below for clarity, as we need to subsequently explain in the following section how the modifications need to be done for the dirty (partially labelled) data in the ConGen model:

a. VAE cost function for completely labeled entries in the minibatch (**Equation 1**):

$$\mathcal{L}(x, y) = -\sum_{i=1}^{n_L}\sum_{j=1}^{n_x}\left(x_{i,j}\ln x_{D,i,j} + (1-x_{i,j})\ln(1-x_{D,i,j})\right)$$

$$+ \sum_{i=1}^{n_L}\frac{1}{2}\left(n_y \ln 2\pi + \ln(det(C)) + \sum_{j=1}^{n_y}(y_{L,i,j} - E_j)\sum_{k=1}^{n_y}(y_{L,i,k} - E_k)C_{k,j}^{-1}\right)$$

$$- \sum_{i=1}^{n_L}\sum_{j=1}^{n_z}\frac{1}{2}\left(1 + \ln\sigma(z_{i,j})^2 - \mu(z_{i,j})^2 - \sigma(z_{i,j})^2\right)$$

b. VAE cost function for completely unlabeled entries in the minibatch (**Equation 2**):

$$\mathcal{U}(x) = -\sum_{i=1}^{n_L}\sum_{j=1}^{n_x}\left(x_{i,j}\ln x_{D,i,j} + (1-x_{i,j})\ln(1-x_{D,i,j})\right)$$

$$+ \sum_{i=1}^{n_L}\frac{1}{2}\left(\sum_{j=1}^{n_y}C_{j,j}^{-1}\sigma(y_{P,i,j})^2 + \sum_{j=1}^{n_y}(y_{P,i,j} - E_j)\sum_{k=1}^{n_y}(y_{P,i,k} - E_k)C_{k,j}^{-1} - n_y + \ln(det(C)) - \sum_{j=1}^{n_y}\ln\sigma(y_{P,i,j})^2\right)$$

$$- \sum_{i=1}^{n_L}\sum_{j=1}^{n_z}\frac{1}{2}\left(1 + \ln\sigma(z_{i,j})^2 - \mu(z_{i,j})^2 - \sigma(z_{i,j})^2\right)$$

c. Regression cost function for completely labeled entries (**Equation 3**):

$$\mathcal{R}_{SSVAE}(x,y) = \beta \sum_{i=1}^{n_L}\sum_{j=1}^{n_y}\left(y_{L,i,j} - \mu(y_{P,i,j})\right)^2$$

where $C = Cov(y_L)$ and $E = E(y_L)$ are the label covariance matrix and mean values constructed from the entire fully labelled training set, $\mu$ is the mean function, $\sigma$ is the standard deviation function, $\beta$ is the tradeoff hyperparameter between generative and supervised learning, while $n_L, n_U, n_x, n_y,$ and $n_z$ represent the number of minibatch's completely labeled entries, completely unlabeled entries, and dimensions of $x, y,$ and $z$ respectively. Finally, the total minibatch cost function is simply $Cost_{SSVAE} = \mathcal{L} + \mathcal{U} + \mathcal{R}_{SSVAE}$.

Finally, once the training is finished, the decoder sub-model can be extracted and be run independently by specifying the conditional property input $y$ and the randomly sampled latent space input $z$ to conditionally generate the desired molecule outputs, where a beam search algorithm is used for converting output $x_D$ to a molecule SMILES. The primary disadvantage of this approach is that the training dataset must be either fully labelled or fully unlabelled. The reason the SSVAE model splits the problem as specified in **Figure 1** above is because it simplifies the model dataflow, math, and behaviour tremendously. In practice, training datasets of interest likely consist of molecules with incomplete labels, in addition to the completely labelled or unlabelled molecules. This is especially so, if the training molecule database is either taken from a publicly available database (like PubChem experimental data[10]) or combined from several different databases. Neither of these practical types of "dirty" datasets will work for training the baseline SSVAE model, thereby severely limiting the type of conditional molecule generation which can be done, especially when multi-property conditional molecule generation is desired. This is typically the case for battery electrolyte or pharmaceutical drug molecule virtual screening.

## Enabling Sparse Labelled Data Utilization using ConGen Model

We modify the SSVAE model into the ConGen model, which is explicitly designed to work with "dirty" training data, thereby enabling the usage of significantly larger number of training data sources including those merged from different public and private sources. This enables us to perform conditional molecule generation tasks which are previously not possible using the SSVAE model. For example, given a large labeled molecule dataset from ZINC[11] (containing *Mol.Wt*, hydrophobicity *LogP*, and drug-likeness *QED*) and another similarly large molecule dataset from Materials Project Electrolyte Genome[12] (containing *Mol.Wt*, *EA*, and *IP*), we can train a conditional generative model which can generate molecules with multiple simultaneous constraints on the *Mol.Wt*, *LogP*, and *EA* values (known useful properties for screening lithium battery electrolytes). Given these diverse sources of training data, the original SSVAE model cannot be trained on the combined database of *Mol.Wt*, *LogP*, and *EA* labels because the training data label is sparse. ConGen on the other hand has no such limitation, allowing users to mix non-ideal practical data from multiple sources as desired.

The primary idea of the ConGen model is to take the general high-level architecture of the SSVAE model, but then modify all its components as needed in order to enable the usage of dirty training data. We have re-written the entire SSVAE model from the original TensorFlow 1.0 version into a PyTorch version to enable better model flexibility, before further implementing the necessary modifications to enable the usage of sparse training data labels. When this PyTorch version is trained on the original SSVAE training data (only fully labelled and fully unlabelled molecules) using the same hyperparameter training settings ($n_{trn}$ = 285k training molecules with 50:50 labelled/unlabelled molecule split, $n_{val}$ = 15k validation molecules, $n_{tst}$ = 10k test molecules, $\beta = 10^4$, Adam optimizer learning rate $LR = 10^{-4}$), we obtain accuracy metrics for property prediction, unconditional and single-property conditional molecule generation tasks (only *Mol.Wt* = 250 Da constraint is used, because the original SSVAE code only allows single-property constraint) equivalent to the TensorFlow version (**Table 1**). 100 molecules are generated on both unconditional & conditional generation tasks.

| Task | Property | SSVAE | ConGen |
|---|---|---|---|
| Predictor Regression MAE | Mol.Wt (Da) | 0.95 | 1.22 |
| | LogP | 0.06 | 0.08 |
| | QED | 0.013 | 0.014 |
| Decoder Unconditional Generation | Mol.Wt (Da) | 360 ± 65 | 363 ± 64 |
| | LogP | 2.95 ± 1.06 | 3.01 ± 1.07 |
| | QED | 0.723 ± 0.142 | 0.713 ± 0.154 |
| Decoder Conditional Generation | Mol.Wt (Da) | 249 ± 6 | 251 ± 5 |
| | LogP | 2.38 ± 0.89 | 2.13 ± 0.91 |
| | QED | 0.810 ± 0.072 | 0.816 ± 0.095 |

**Table 1 | Comparison between SSVAE (TensorFlow 1.0) and baseline ConGen (PyTorch) model on the original SSVAE model tasks.** The baseline ConGen is equivalent to SSVAE, except that it is implemented in PyTorch. This comparison is performed on SSVAE 'clean' original training dataset, which only contains fully labelled and fully unlabelled molecules. Identical training hyperparameter settings are used, and relatively equivalent performance metrics are obtained. The slight differences can be attributed to the high aggressivity of the original model's training hyperparameter settings. For the property prediction task, predictor sub-model is utilized to calculate mean absolute error (MAE) with respect to the training labels. For the unconditional / conditional generation tasks, the decoder sub-model is used to generate the molecules and the molecules property labels are calculated using RDKit cheminformatics library.

Once we have confirmed that the two models are equivalent, the input data preprocessing and molecule data workflow inside the ConGen sub-models are modified (**Figure 2**). First, we enable the ability to merge molecule training data labels with different types of property labels into a new property label matrix $y_L$. This will cause a significant fraction of the merged database to contain missing [molecule, property] entry labels. For entries with no label available from all the databases, we designate the property label as invalid. This can be done by generating a mask matrix $M$ containing '0' for invalid entries and '1' for entries with available property values. For entries where multiple property labels are available from different databases, we choose the available label from the latest database being merged. Both $y_L$ and $M$ matrices are now required as inputs into the ConGen model. ConGen no longer differentiates data workflow based on whether the molecule is fully labelled or fully unlabelled. ConGen instead implements a selector for the intermediate label $y$ which choose whether to utilize existing label $y_L$ or the predicted property label $y_P$ generated by the predictor sub-model depending on the value of the mask $M$ (**Equation 4**):

$$y(i,j) = \begin{cases} y_L(i,j) & if \ M(i,j) == 1 \\ y_P(i,j) & if \ M(i,j) == 0 \end{cases}$$

where $i$ and $j$ denote the molecule and property type indices, respectively. With this modification, a unified data workflow can be utilized for fully labelled, fully unlabelled, and partially unlabelled molecules. Furthermore, when the molecule in the minibatch is either fully labelled or fully unlabelled the mathematical operations performed on them within the ConGen model will be identical to those performed in the SSVAE model.

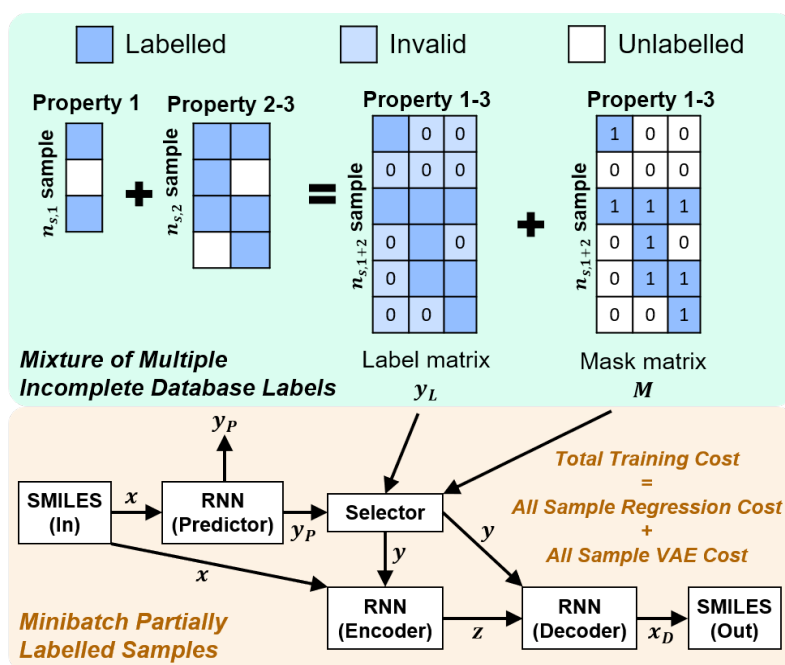

**Figure 2 | Dirty training label data merging and high-level dirty data workflow within the ConGen model.** ConGen model no longer differentiates between fully labelled, fully unlabelled, and partially labelled molecule inputs. The unified data workflow is controlled by the mask matrix $M$. $n_{s,1}$, $n_{s,2}$, and $n_{s,1+2}$ denote the number of samples within the first, the second, and the merged property databases respectively.

However, it is not as straightforward with respect to the training cost function and subsequent molecule generation. It is important to recognize that the implementation of the training cost function within the SSVAE model is heavily dependent on whether the molecule is fully labelled or fully unlabelled. The SSVAE cost function consists of three major elements, designed to ensure that the predictor, encoder, and decoder are all accurate (**Equation 1-3**) and we need to design the dirty data VAE cost function substitute for $\mathcal{L}$ and $\mathcal{U}$ because we no longer have fully labelled and fully unlabelled molecules. It is worth noting that during the execution of the original SSVAE model, there is no interaction between molecule inputs within a minibatch (e.g. if molecule A and B are processed simultaneously, the model output $x_D$ for both molecules are not influenced by the fact that the other molecule is also simultaneously processed. This ensures that any intermediate values for a molecule ($y_L$, $y_P$, $z$, $x_D$, etc) are solely determined by that molecule input $x$. Because of this, the implementation of a new cost function for the ConGen model becomes less complicated. There is a significant overlap of terms between $\mathcal{L}$ and $\mathcal{U}$, enabling us to design a new VAE cost function $\mathcal{G}$ for the ConGen model which takes partially labeled entries utilizing our mask matrix $M$ (**Equation 5**). When the entries are all completely labeled, the entries of $M$ will all be 1, and $\mathcal{G}$ should be converted to $\mathcal{L}$, except for some constant terms that do not affect the training. When the entries are all completely unlabeled, the entries of $M$ will all be 0, and $\mathcal{G}$ should be converted to $\mathcal{U}$, again, except for some constant terms. Similarly, our new regression cost function $\mathcal{R}_{ConGen}$ should only sum over labeled entries in the minibatch. By ensuring this behavior, the subsequent ConGen cost function differentiation and model parameter optimization will work exactly like the SSVAE versions when completely labeled / unlabeled data are supplied. However, it will also now work for dirty sparsely labelled training data. Henceforth, we define new cost functions for the ConGen minibatch, especially meant for dirty data:

   a. VAE cost function for dirty labeled entries in the minibatch (**Equation 5**):

$$\mathcal{G}(x,y) = -\sum_{i=1}^{n_S}\sum_{j=1}^{n_x}\left(x_{i,j}\ln x_{D,i,j} + (1-x_{i,j})\ln(1-x_{D,i,j})\right)$$

$$+ \sum_{i=1}^{n_S}\frac{1}{2}\left(n_y \ln 2\pi + \sum_{j=1}^{n_y} M_{i,j}(y_{i,j}-E_j)\sum_{k=1}^{n_y} M_{i,k}(y_{i,k}-E_k)C_{k,j}^{-1}\right)$$

$$+\sum_{i=1}^{n_S}\frac{1}{2}\left(\sum_{j=1}^{n_y}C_{j,j}^{-1}(1-M_{i,j})\sigma(y_{i,j})^2 + \sum_{j=1}^{n_y}(1-M_{i,j})(y_{i,j}-E_j)\sum_{k=1}^{n_y}(1-M_{i,k})(y_{i,k}-E_k)C_{k,j}^{-1} - n_y - \sum_{j=1}^{n_y}(1-M_{i,j})\ln\sigma(y_{i,j})^2\right)$$

$$-\sum_{i=1}^{n_S}\sum_{j=1}^{n_z}\frac{1}{2}\left(1+\ln\sigma(z_{i,j})^2 - \mu(z_{i,j})^2 - \sigma(z_{i,j})^2\right)$$

   b. Regression cost function for dirty labeled entries in the minibatch (**Equation 6**):

$$\mathcal{R}_{ConGen}(x,y) = \beta\sum_{i=1}^{n_S}\sum_{j=1}^{n_y} M_{i,j}\left(y_{i,j} - \mu(y_{P,i,j})\right)^2$$

where $n_S$ refers to the number of all samples in the dirty data minibatch. It is straightforward to prove that under this scheme, $\mathcal{G}$ is converted to either $\mathcal{L}$ or $\mathcal{U}$ depending on the values of $M$, except for constant terms which do not have any impact on the model parameter optimization process. Note that, compared to the SSVAE cost functions, we have intentionally removed the constant terms $\ln(det(C))$ from the cost function above for numerical reasons we will describe in the following paragraph related to the dirty data covariance matrix $C$. Crucially, under this new cost function only the corresponding labeled / unlabeled matrix elements from $\mathcal{L}$ and $\mathcal{U}$ contributes to the summation over $n_s$ and $n_y$ forming $\mathcal{G}$. The total minibatch cost function is now simply $Cost_{ConGen} = \mathcal{G} + \mathcal{R}_{ConGen}$.

It is important to note that because we only have partially labeled entries, we do not have complete entries for $\mathbf{y_L}$ and correspondingly $C = Cov(\mathbf{y_L})$ and $E = E(\mathbf{y_L})$ can only be calculated using the incomplete entries, making these matrices ill-defined especially $Cov(\mathbf{y_L})$. For an SSVAE model, $C$ is well-defined because it is straightforward to completely discard the unlabelled molecule entries from the training set and calculate $C$ and $E$ directly from the completely labelled molecule entries (this will be a good approximation as long as there is a large number of fully labelled molecules which is a good chemical representation of the full training dataset). This can be done once during the model construction and be set at those values throughout the entire model training. However, this strategy does not work for ConGen because the training data is dirty. In this case, it only makes sense to calculate the label mean $E$ from the valid entries and ignore the invalid values in the $\mathbf{y_L}$ matrix. Similarly, it makes more sense to calculate covariance matrix $C$ entries from the available $\mathbf{y_L}$ matrix entries while ignoring the invalid entries. In other words, we have the following situation for $E$ and $C$ calculation (**Equation 7-8**):

$$E_j = E(\mathbf{y_L})_j = \frac{\sum_{i=1}^{n_S} \mathbf{y}_{L,i,j} \mathbf{M}_{i,j}}{\sum_{i=1}^{n_S} \mathbf{M}_{i,j}}$$

$$\mathbf{C}_{j,k} = Cov(\mathbf{y_L})_{j,k} = \frac{\sum_{i=1}^{n_S}(\mathbf{y}_{L,i,j} - E_j)(\mathbf{y}_{L,i,k} - E_k)\mathbf{M}_{i,j}\mathbf{M}_{i,k}}{\left(\sum_{i=1}^{n_S} \mathbf{M}_{i,j}\mathbf{M}_{i,k}\right) - 1}$$

In a clean training data like the ones being used in the SSVAE model, all entries of the mask matrix $M$ are all 1's, and it can then mathematically be proven that the covariance matrix $C$ will always be a positive semi-definite (PSD) matrix. Correspondingly, in SSVAE the log-determinant term $\ln(det(C))$ in the cost function above will always be well-defined. The mathematical guarantee breaks down when the entries of mask matrix $M$ are no longer all 1's, however.[13] Consequently, we can get training errors due to attempting log operations on negative numbers. Nevertheless, because the term $\ln(det(C))$ is just a constant, we can remove it from the ConGen cost function without any mathematical training consequences as we have done in **Equation 5**.

The real physical issue arises from the quality of $E$ and $C$ themselves. When we have low availability of training data label (a lot of 0 entries in the mask matrix $M$), we will have significant problems because the $E$ and $C$ matrices do not accurately represent the real molecule property labels, especially when we have many invalid labels in the training dataset. Keeping the values of $E$ and $C$ the same throughout the training iterations mean we will have poor control on the conditionally generated molecules' properties after subsequent model training and conditional generation processes. We can mitigate this problem by using imputation technique[13] to re-calculate $E$ and $C$ using predicted molecule property labels from the predictor sub-model when there is no valid label in $y_L$. In other words, we track minibatch $y$ from the selector (**Equation 4**, **Figure 2**) throughout a training epoch, and re-calculate $E$ and $C$ using $y$ instead of using $y_L$ after each training and validation cycle in the epoch has been completed. This update is performed iteratively throughout the training, and it is important to store the final $E$ and $C$ as part of the ConGen model parameter because subsequent molecule generation tasks need to be performed using these higher quality $E$ and $C$ parameters (**Equation 8-9**):

$$E_j = E(y)_j = \frac{\sum_{i=1}^{n_a} y_{i,j}}{n_a - 1}$$

$$C_{j,k} = Cov(y)_{j,k} = \frac{\sum_{i=1}^{n_a}(y_{i,j} - E_j)(y_{i,k} - E_k)}{n_a - 1}$$

where $n_a$ is the number of all molecules in the training dataset. The quality of $E$ and $C$ are not very good in the beginning of the training. However, as the predictor sub-model gets more accurate during subsequent training iterations, $E$ and $C$ will represent the real sample population better and we correspondingly achieve better molecule property prediction and conditional generation accuracy in the end.

We also take advantage of the modular nature of the ConGen model (inherited from the modularity of SSVAE) to further improve model performance on dataset with rare training property labels (such as ab-initio simulation or experimental properties). It is straightforward to implement transfer learning in ConGen by replacing the recurrent neural networks (RNN) in the predictor and encoder sub-models with a bidirectional encoder representations from transformer (BERT) model pre-trained on a much larger (but 'cheaper') molecule property dataset. Here we use the ChemBERTa model, which is a large-scale self-supervised transformer-based pretraining model which only requires molecule SMILES as input and has been thoroughly evaluated.[14] During the sub-model construction, we add a fully connected network linear layer on top of the transferred ChemBERTa model. We nickname this type of transferred model 'BERT' from here onward. When BERT is used to substitute the RNN encoder, the entire ChemBERTa layers' parameters are frozen. However, when BERT is used to substitute the RNN predictor, the last ChemBERTa layer's parameters can be fine-tuned by the PyTorch optimizer. While we do not substitute the RNN decoder with other types of decoder sub-model, in principle it is straightforward to do so as well if desired. For the standard ConGen model training with just RNN sub-models, we set the Adam optimizer $LR = 10^{-4}$ and clip the gradients absolute value to a maximum of $10^2$. For the ConGen model training with BERT predictor and decoder sub-model substitutions, we have significantly lower Adam optimizer $LR = 3 \times 10^{-5}$ for the BERT-based sub-models, and $LR = 10^{-3}$ is used for optimizing the decoder sub-model parameters.

Finally, we demonstrate the resulting capability of the ConGen model on dirty dataset in **Table 2**. The training data labels are mixed from two different databases: 1) ZINC database containing properties such as *Mol.Wt*, *LogP*, and *QED*[11] used in the SSVAE publication,[8] 2) Materials Project

Electrolyte Genome database containing properties such as *IE* and *EA*.[12] The ConGen model is trained on all 5 of these properties, which cannot be done by the SSVAE model. As an example of multi-property conditional generation, we query the models to generate molecules with 3 simultaneous properties: *Mol.Wt* = 250 Da, *LogP* = 2.5, and *IE* = 5 eV. The corresponding regression and conditional generation results are given below in **Table 2**. We validate the properties of the generated molecules using RDKit[15] (for *Mol.Wt* and *LogP*) and quantum chemistry (for *IE*, see **Methods**). We see that overall, the BERT-based ConGen has worse performance than the RNN-based ConGen model on property prediction tasks, but is relatively equivalent to the RNN-based ConGen on conditional generation tasks (good on *Mol.Wt* and *LogP*, but less accurate on *IE*). We have expected the transferred BERT-based ConGen to perform worse than the RNN-based ConGen on abundant property label such as *Mol.Wt* and *LogP* and better than RNN-based ConGen on rare property label such as *IE*. The fact that both RNN and BERT-based ConGen shows relatively equivalent performance for molecular conditional generation tasks merits further future investigation. We hypothesize that we still have insufficient number of quantum chemistry property training labels from just the Materials Project Electrolyte Genome database,[12] and that a more accurate and data-efficient predictor sub-model is still needed. Currently the BERT-based ConGen is computationally more expensive while offering no significant improvement over the RNN-based ConGen, so we focus solely on using RNN-based ConGen in the following large-scale electrolyte diluent screening section.

| Task | Model | *Mol.Wt* (Da) | *LogP* | *QED* | *EA* (eV) | *IE* (eV) |
|---|---|---|---|---|---|---|
| Predictor Regression Test Set *MAE* | RNN | 2.70 | 0.05 | 0.009 | 0.20 | 0.16 |
| | BERT | 6.07 | 0.15 | 0.017 | 0.22 | 0.19 |
| Decoder Unconditional Generation | RNN | 312 ± 95 | 2.07 ± 1.28 | 0.677 ± 0.124 | 1.79 ± 0.84 | 5.99 ± 0.44 |
| | BERT | 271 ± 145 | 2.15 ± 1.11 | 0.583 ± 0.138 | 1.72 ± 0.82 | 6.40 ± 0.34 |
| Decoder Conditional Generation | RNN | 248 ± 4 | 2.55 ± 0.23 | 0.672 ± 0.082 | 2.06 ± 0.55 | 6.53 ± 0.62 |
| | BERT | 252 ± 3 | 2.45 ± 0.36 | 0.756 ± 0.127 | 1.80 ± 0.64 | 6.36 ± 0.41 |

**Table 2 | ConGen model performance comparison on 'dirty data' tasks, including both RNN-based ConGen and BERT-based ConGen.** This 'dirty data' task cannot be done with the original SSVAE model but is useful in practice for conditional generative model training because molecule property labels are often unavailable or incomplete. Including a pre-trained BERT can increase the predictor sub-model's ability on 'rare' properties such as *EA* and *IE*, even though in some cases it may reduce the predictor sub-model's performance on 'common' properties (*Mol.Wt*, *LogP*, and *QED* in this case). The conditional generation co-constraints are *Mol.Wt* = 250 Da, *LogP* = 2.5, and *IE* = 5 eV. Regression *MAE* is calculated using property labels from the database, while generated molecules' properties are validated using either RDKit library or *ab-initio* simulation.

# Use Case Example: Lithium-Ion Battery Localized High Concentration Electrolyte Diluent Screening

Finally, we demonstrate the usage of the ConGen model on a practical example: generating the chemical space for further virtual screening of Li-ion battery localized high concentration electrolyte (LHCE) diluent molecules. Recent progress in the development of Li-ion battery electrolytes have led to the discovery of LHCE-type of electrolytes, which microscopically look like that of high salt concentration electrolyte (HCE), but macroscopically look more like a conventional electrolyte.[16] The LHCE is useful because it is stable over a wide electrochemical window, in addition to forming stable solid electrolyte interphase (SEI) layer which is important for the long-term stability of the battery.[17,18] From a cost perspective, the LHCE is also important because it can reduce the required amount of Li-salt used, versus that of HCE which requires a large amount of expensive Li-salt.[19] Finally, LHCE can have significantly lower solution viscosity than HCE, which is useful not just for improving the electrolyte's lithium ion transport properties, but also for enabling better electrode wetting which helps to better optimize the energy capacity of Li-ion battery cells.[20] Chemically, what differentiates LHCE from HCE and conventional electrolytes is the addition of small molecules which act as a diluent in the electrolyte.[16] These diluent molecules are typically hydrofluoroether (HFE) such as bis(2,2,2-trifluoroethyl) ether (BTFE) and 1,1,2,2-tetrafluoroethyl-2,2,3,3-tetrafluoropropyl ether (TTE), or fluorinated orthoformate such as tris(2,2,2-trifluoroethyl) orthoformate (TFEO).[16–18] The unique trait of these types of compound is that while they are sufficiently polar, they are less polar than the Li-salt anions being used in the LHCE. Consequently, at the right concentration range the Li$^+$ cations will primarily coordinate with the polar salt anions in the first Li$^+$ solvation shell. The diluent molecules will then mostly coordinate with these salt clusters from the second shell onward using their polar oxygen atoms. Furthermore, the fluorinated components of the diluent molecules will tend to form their own non-polar network in the LHCE. Consequently, the addition of diluents into LHCE ensures that locally the salt cluster looks like that of HCE (more stable), while macroscopically the diluents separate these salt clusters and ensure that the solution is less viscous, ionically conductive, and ideally inflammable (due to the proportion reduction of flammable solvent molecules in LHCE).

Many criteria need to be satisfied by these LHCE diluent molecules such as electrochemical stability, inflammability, and low viscosity. While there are several known working LHCE diluents, it is important to find more relevant compounds in this field to enrich the diluent chemical space suitable for the optimization of specific types of Li-ion batteries. We apply the ConGen model to generate candidate molecules for LHCE diluents through structural chemical properties such as: *Mol.Wt*, $n_F$, $n_O$, *IE*, *EA*, and *Log.Vis*. To achieve this, we train ConGen model on a mixture of 5 publicly available datasets:

- *Mol.Wt* database from ZINC[8,11] (310,000 unique entries)
- *Mol.Wt*, simulated *IE*, *EA* database from the Materials Project Electrolyte Genome[12] (62,274 unique entries)
- *Mol.Wt*, simulated *IE*, *EA* database from Austin Apple Github[21] (26,394 unique entries)
- Oxyfluorocarbon $n_F$, $n_O$ database from PubChem[10] (200,000 unique entries)
- Experimental *Log.Vis* database from literature[22] (322 unique entries)

Where applicable, each of these databases are supplemented with the corresponding molecule *Mol.Wt*, $n_F$, and $n_O$ missing property labels because it is computationally efficient and inexpensive to do so using RDKit.[15] The combined database has 571,023 unique molecules. Finally, we evaluate the model's performance. Based on known existing LHCE diluents, we hypothesize that we need the following properties for the LHCE diluent molecules:

- Electrochemical properties:   *EA* <= 0.5 eV, *IE* >= 7.0 eV
- Viscosity property:   *Log.Vis* <= 0.0
- Structural properties:   *Mol.Wt* <= 300, $n_F$ >= 4, $n_O$ = 1-2

Within the framework of ConGen, we can implement this multi-condition molecular structure generation task by simply deploying simultaneous property label 'anchors' as the decoder input during the generation cycle. For example, we may choose the following label anchors to satisfy the conditions stated above:

1. *EA* = 0 **or** 0.5 eV
2. *IE* = 7.0 **or** 7.5 eV
3. *Log.Vis* = -0.1 **or** 0.0
4. *Mol.Wt* = 250 **or** 300 Da.
5. $n_F$ = 4 **or** 6
6. $n_O$ = 1 **or** 2

We correspondingly have $2^6$ = 64 combinations of multi-constraint property anchors we can use for the conditional generation in the example above. For each set of anchors, we generate 5 molecule samples resulting in 320 conditionally sampled molecules using our RNN-based ConGen model (**Query 1**). The training data label distributions, based on just available property labels, is shown below in **Figure 3**.

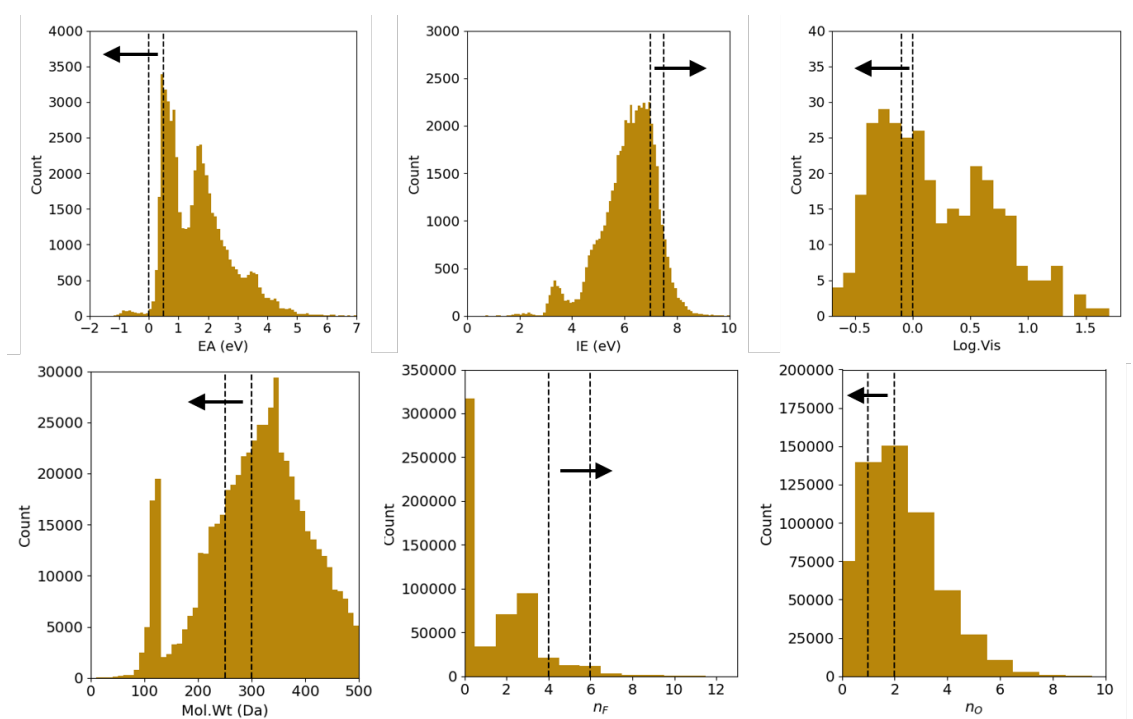

**Figure 3 | Training data molecular property label distribution**. The dashed lines indicate the property label 'anchors' we will use for subsequent conditional molecular generation. The arrows indicate the preferred generated molecules' property range. The anchors are respectively: *EA* = [0,0.5], *IE* = [7.0,7.5], *Log.Vis* = [-0.1,0.0], *Mol.Wt* = [250,300], $n_F$ = [4,6], $n_O$ = [1,2].

Regression on the test set, unconditional molecule generation, as well as conditional molecule generation results are shown below in **Figure 4** and **Table 3**. In order to calculate the ground truth property label values for the generated molecules, several methods are employed. For *Mol.Wt*, $n_F$, and $n_O$, simple cheminformatics tool such as RDKit can be used to quickly calculate their true values. For *EA* and *IE*, we used quantum chemistry calculations with identical calculation settings to the prior work[12] to calculate the true values. We see that we have excellent control over the generated molecules' structural properties (*Mol.Wt*, $n_F$, and $n_O$) and *IE*, although we observe a positive shift of approximately 2.0 eV on the generated molecules' *EA* compared to the mean of the anchors' *EA* (0.25 eV). We hypothesize that this systemic shift may be caused by the slight difference in our adiabatic EA calculation workflow compared to the procedure utilized by the Materials Project Electrolyte Genome team, as well as the fact that we query the ConGen model to generate molecules with *EA* label anchors at the extreme left end of the training dataset *EA* label distribution (making this the most difficult constraint out of the 6 co-constraints we have employed).

We currently have no experimental validation capability to measure *Log.Vis* for the generated molecules, so unfortunately no accuracy metric can be displayed for these molecules' *Log.Vis* property. Nevertheless, we have listed all the molecules that the ConGen model has generated based on their property label input anchors in **Table 4** for future validation by other research groups with experimental capabilities. Additional molecular property criteria are likely needed to further improve the quality of the generated LHCE diluent candidate molecules. Inclusion of further molecular property constraints to help refine this generated LHCE diluent chemical space further should be straightforward, as it can be done by simply adding a new comma-separated-value (CSV) file containing the desired molecular properties for training. Out of the 320 generated molecule SMILES, 6 are invalid molecules, 3 are duplicates, and 5 are within the training set. We have correspondingly generated 306 new unique candidate molecules from this query for computational validation purposes. We further generate 64,000 candidate molecules using the RNN-based ConGen model (1,000 queries for each of the anchor combinations, see **Figure 4**) although neither *EA* nor *IE* ab-initio computational validation is done for these additional molecules due to the high computation costs (**Query 2**). Out of this new query for 64,000 molecules, 1,486 are invalid, 41,117 are duplicates, and 356 are within the training set. Correspondingly, **Query 2** generates 21,041 new unique candidate LHCE diluent molecules. Future work is needed to reduce the number of large-scale-query duplicates.

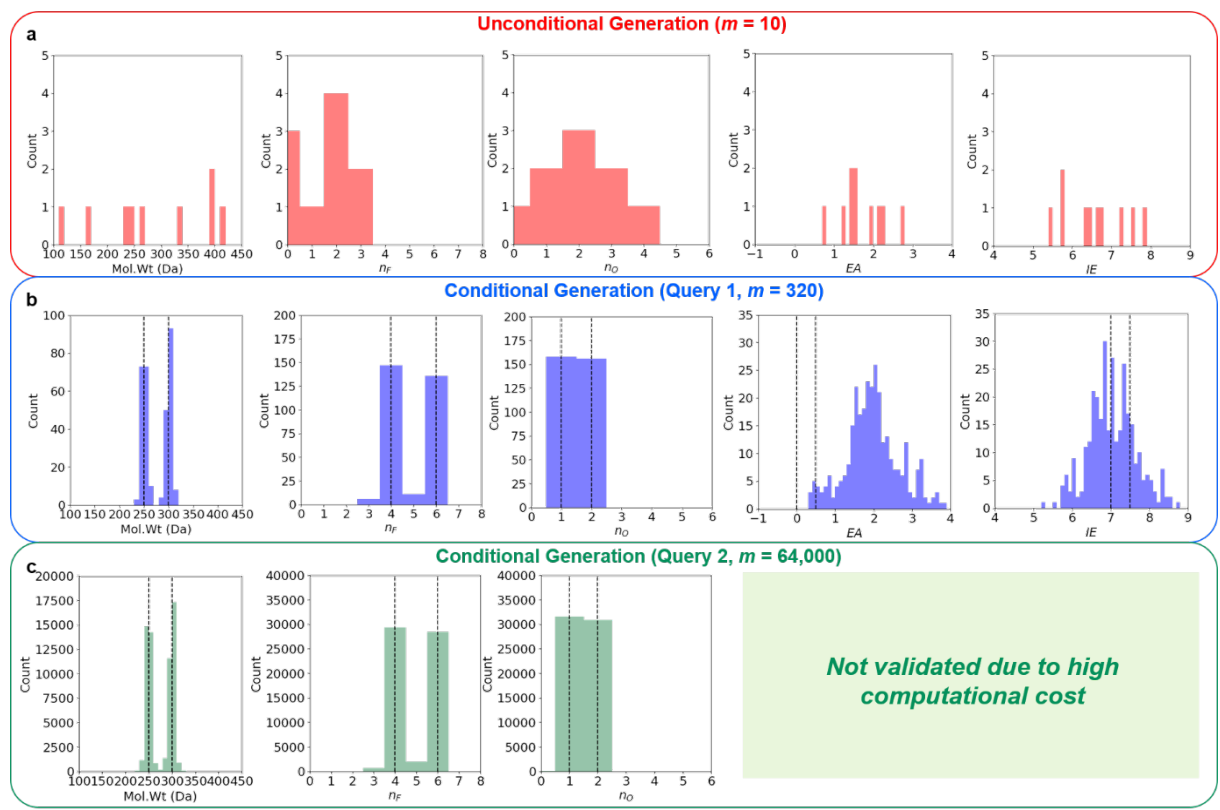

**Figure 4 | ConGen unconditional and multi-constraint conditional molecular generation property distributions**. **(a)** Unconditional molecule generation showing property distribution without constraints. When multiple co-constraints are utilized for the conditional generation, we have very targeted molecule generation. Structural and electrochemical stability properties validation for **Query 1** with 320 molecules is shown in **(b)**, while structural property validation for **Query 2** with 64k molecules is shown in **(c)**. We can see that the molecules generated with simultaneous multi-property constraints still obey their conditional property anchors quite well (simultaneously, although the hardest property *EA* distribution is slightly shifted), and that the generated molecules' property distribution is very different from molecules generated with no property constraint. Conditional generation property anchor inputs are shown as dashed lines.

| Task | *Mol.Wt* (Da) | $n_F$ | $n_O$ | *EA* (eV) | *IE* (eV) | *Log.Vis* |
|---|---|---|---|---|---|---|
| Predictor Regression Test Set *MAE* | 1.60 | 0.01 | 0.02 | 0.20 | 0.21 | 0.14 |
| Decoder Unconditional Generation | 302 ± 110 | 1.50 ± 1.12 | 2.30 ± 1.42 | 1.71 ± 0.54 | 6.58 ± 0.75 | N/A |
| Decoder Conditional Generation (Query 1) | 275 ± 26 | 5.02 ± 1.08 | 1.50 ± 0.50 | 1.99 ± 0.73 | 7.04 ± 0.61 | N/A |
| Decoder Conditional Generation (Query 2) | 274 ± 26 | 5.02 ± 1.05 | 1.49 ± 0.50 | N/A | N/A | N/A |

**Table 3 | Molecular property prediction accuracy and the generated molecule's property distribution statistics for LHCE diluent molecules**. From the regression test result, we can see the predictor sub-model is reasonably accurate in predicting molecular property. In addition to that, the discrepancy in distributed molecules' properties for unconditional vs conditional

generation cases show that the conditional generator is generating the right molecules, based on the property label input anchors we have chosen. Regression *MAE* is calculated using property labels from the database, while generated molecules' properties are validated using either RDKit library or *ab-initio* simulation.

**Table 4 | Candidate Li-ion battery LHCE diluent molecules generated with multi-constraint ConGen model (Query 1).**

| ['EA', 'IE', 'LogVis', 'MolWt', 'n_F', 'n_O'] : [0.5, 7.0, 0.0, 250, 4.0, 1.0] | ['EA', 'IE', 'LogVis', 'MolWt', 'n_F', 'n_O'] : [0.0, 7.0, 0.0, 250, 4.0, 1.0] |
|---|---|
| Cc1ccc(OCC(F)(F)C(F)(F)Cl)cc1<br>Nc1c(F)cc(OC(F)(F)F)cc1CCl<br>CC(C)C(=O)Nc1ccc(F)c(C(F)(F)F)c1<br>CN(C)C(=O)Nc1c(F)c(F)cc(F)c1CF<br>OCc1nc(C(F)(F)F)nc2c(F)cccc12 | CN(C)C(=O)Nc1cc(F)cc(C(F)(F)F)c1<br>CC(C)(C)OC(c1cc(F)cc(F)c1)C(F)F<br>CCOC(=N)c1c(F)cccc1CC(F)(F)F<br>Nc1cc(C(F)(F)F)ccc1OCCC(F)C<br>Fc1cc(OC(F)(F)F)cc(C2CCNC2)c1 |
| ['EA', 'IE', 'LogVis', 'MolWt', 'n_F', 'n_O'] : [0.5, 7.0, 0.0, 250, 4.0, 2.0] | ['EA', 'IE', 'LogVis', 'MolWt', 'n_F', 'n_O'] : [0.0, 7.0, 0.0, 250, 4.0, 2.0] |
| CN(C)C(=O)OC1(C(F)(F)F)CCC(F)C1<br>CCOC(=O)c1cc(C(F)(F)F)nc(F)c1C<br>CC(=O)NCC(O)c1c(F)c(F)cc(F)c1F<br>OCc1c(OC(F)(F)F)ccc(F)c1C1CC1<br>COc1ccc(OCC(F)(F)C(F)F)c(C)c1 | Oc1cc(OC(F)(F)F)cc(F)c1CCl<br>OC(OCC(F)(F)C(F)F)c1ccsc1<br>COC(=O)CC(CC(F)(F)F)c1cccccc1<br>Cc1cc(OCC(F)(F)C(F)(F)CO)ccn1<br>COC(=O)Cc1c(F)cnc(C(F)F)c1CF |
| ['EA', 'IE', 'LogVis', 'MolWt', 'n_F', 'n_O'] : [0.5, 7.0, 0.0, 250, 6.0, 1.0] | ['EA', 'IE', 'LogVis', 'MolWt', 'n_F', 'n_O'] : [0.0, 7.0, 0.0, 250, 6.0, 1.0] |
| Oc1nc(F)cc(CC(F)(F)C(F)(F)F)n1<br>FC(C(F)(F)F)C(F)(F)COC1CCCC1<br>FC(F)(F)C(F)(F)COc1cccccc1F<br>OC(CCC(F)(C(F)(F)F)C(F)(F)C1CC1<br>Cn1nc(C(F)(F)F)c(C(F)(F)F)c1CO | CC(OC(F)(F)F)c1cccccc1C(F)(F)F<br>Cc1ncc(OC(F)(F)F)nc1C(F)(F)F<br>OC(CCC(F)(F)C(F)(F)F)C1CC1<br>Fc1ccc(OCC(F)(F)F)cc1C(F)F<br>NC(c1ccoc1)C(C(F)(F)F)C(F)(F)F |
| ['EA', 'IE', 'LogVis', 'MolWt', 'n_F', 'n_O'] : [0.5, 7.0, 0.0, 250, 6.0, 2.0] | ['EA', 'IE', 'LogVis', 'MolWt', 'n_F', 'n_O'] : [0.0, 7.0, 0.0, 250, 6.0, 2.0] |
| COC(=O)CCCC(F)(F)C(F)(F)C(F)F<br>OC(O)c1cc(C(F)(F)F)c(F)c(F)c1F<br>Oc1cc(C(F)(F)F)cc(C(F)(F)F)c1O<br>COC(=O)CCC(F)(C(F)(F)F)C(F)(F)F<br>-- *invalid*-- | OC(O)(CCCC(F)(F)F)CC(F)(F)F<br>CC(=O)NCC(O)(C(F)(F)F)C(F)(F)F<br>CCOc1ccc(C(F)(F)F)c(C(F)(F)F)c1<br>OC(OCC(F)(F)F)c1cccc(F)c1F<br>Oc1c(OC(F)(F)F)cccc1C(F)(F)F |
| ['EA', 'IE', 'LogVis', 'MolWt', 'n_F', 'n_O'] : [0.5, 7.0, 0.0, 300, 4.0, 1.0] | ['EA', 'IE', 'LogVis', 'MolWt', 'n_F', 'n_O'] : [0.0, 7.0, 0.0, 300, 4.0, 1.0] |
| Cc1noc(-c2ccc(C(F)(F)F)cc2Cl)c1C(F)F<br>Cc1cc(OC(F)(F)F)cc(F)c1I<br>COc1cnc(C(F)(F)F)c(F)c1CBr<br>Nc1c(F)cc(Oc2ccc(C(F)(F)F)cc2)c(Cl)c1<br>COc1cnc(-c2ccc(C(F)(F)F)c(F)c2)c(Cl)c1 | Fc1cnc(OC(F)(F)F)c(I)c1<br>O=C(Nc1ncc(C(F)(F)F)cc1Cl)c1cccccc1<br>Cn1cc(-c2noc(-c3cc(F)c(F)c(F)c3F)n2)s1<br>NC(=O)c1c(F)cccc1Nc1cc(C(F)(F)F)ccn1<br>OC(Cc1ccc(C(F)(F)F)c(F)c1)c1cccs1 |
| ['EA', 'IE', 'LogVis', 'MolWt', 'n_F', 'n_O'] : [0.5, 7.0, 0.0, 300, 4.0, 2.0] | ['EA', 'IE', 'LogVis', 'MolWt', 'n_F', 'n_O'] : [0.0, 7.0, 0.0, 300, 4.0, 2.0] |
| OC(Cc1cc(F)c(Br)cc1F)C(F)(F)O<br>COC(=O)CCc1ccc(Cl)cc1C(F)(F)C(F)F<br>CCOC(=O)Nc1c(C(F)(F)F)cc(F)nc1CCl<br>N[C@@H](Cc1cc(F)c(F)c(F)c1F)c1ccc(O)cc1O<br>CCC(=O)NCC(=O)Nc1cc(C(F)(F)F)cc(F)c1C | NC(=O)COc1cccccc1-c1c(F)c(F)cc(F)c1F<br>OC(O)(c1cc(F)c(F)c(F)c1)c1ccc(F)cc1Cl<br>NCc1cnc(OC(F)(F)F)nc1Oc1c(F)cccc1<br>COc1ccc(CNc2cc(F)c(F)c(F)c2F)cc1O<br>FC(F)(F)Oc1cccc(Oc2cc(F)cc(Cl)c2)c1 |

| ['EA', 'IE', 'LogVis', 'MolWt', 'n_F', 'n_O'] : [0.5, 7.0, 0.0, 300, 6.0, 1.0] | ['EA', 'IE', 'LogVis', 'MolWt', 'n_F', 'n_O'] : [0.0, 7.0, 0.0, 300, 6.0, 1.0] |
|---|---|
| COC(c1cc(C(F)(F)F)nc(C(F)(F)F)c1)C1CC1<br>Nc1ccc(C(=O)NCC(F)(F)C(F)(F)C(F)F)cc1<br>NCc1cc(OC(F)(F)F)c(Cl)cc1C(F)(F)F<br>NC(=O)c1cc(C(F)(F)F)cc(C(F)(F)F)c1CCl<br>CC(C)C(=O)Nc1cc(C(F)(F)F)cc(C(F)(F)F)c1 | COc1c(C(F)(F)F)ncc(C(F)(F)F)c1CCl<br>OC(c1ccc(F)cc1)c1c(F)c(F)c(F)c(F)c1F<br>Nc1cc(C(F)(F)F)cc(OC(F)(F)F)c1CCl<br>FC(F)(F)c1cccc(-c2ccc(OC(F)(F)F)cc2)c1<br>Nc1c(F)cccc1Oc1cc(F)c(F)c(C(F)(F)F)c1 |
| ['EA', 'IE', 'LogVis', 'MolWt', 'n_F', 'n_O'] : [0.5, 7.0, 0.0, 300, 6.0, 2.0] | ['EA', 'IE', 'LogVis', 'MolWt', 'n_F', 'n_O'] : [0.0, 7.0, 0.0, 300, 6.0, 2.0] |
| OCc1ccc(C(F)(F)F)cc1OCCC(F)(F)CF<br>OB(O)c1c(C(F)(F)F)ccc(Cl)c1C(F)(F)F<br>COc1ccc(OC(C(F)(F)F)C(F)(F)C(F)F)nc1<br>OCc1c(OC(F)(F)F)ncc(C(F)(F)F)c1C1CC1<br>CS(=O)(=O)Nc1cc(C(F)(F)F)cc(C(F)(F)F)c1 | COc1cc(C(F)(F)F)c(C(F)(F)F)c(CC(N)=O)c1<br>Cc1cc(OC(F)(F)F)nc(OC(F)(F)F)c1CC#N<br>Cc1ccc(COCC(F)(F)C(F)(F)C(F)F)cc1O<br>O=C(O)Cc1cc(C(F)(F)F)nc(C(F)(F)F)c1CN<br>OC(c1cccc(OC(F)(F)F)c1)c1ccc(F)cc1F |
| ['EA', 'IE', 'LogVis', 'MolWt', 'n_F', 'n_O'] : [0.5, 7.0, -0.1, 250, 4.0, 1.0] | ['EA', 'IE', 'LogVis', 'MolWt', 'n_F', 'n_O'] : [0.0, 7.0, -0.1, 250, 4.0, 1.0] |
| Oc1ccc(-c2ccc(F)c(F)c2)c(F)c1F<br>COC(c1c(F)c(F)nc(F)c1F)C1CCC1<br>NCc1cn(CC(F)(F)F)nc1CC(=O)F<br>NCc1cc(OC(F)(F)F)cc(Cl)c1F<br>Cn1cnc(OC(F)(F)F)c1-c1ccc(F)cc1 | Cc1nc(-c2cccc(C(F)(F)F)c2F)c(C)o1<br>CC(NC(=O)C(F)(F)C(F)F)c1ccccc1<br>CC(CO)Nc1ccccc1C(F)(F)C(F)F<br>CC(C)Oc1nc(C(F)(F)F)c(F)cc1CN<br>OCC1Cc2cc(C(F)(F)F)cc(F)c2S1 |
| ['EA', 'IE', 'LogVis', 'MolWt', 'n_F', 'n_O'] : [0.5, 7.0, -0.1, 250, 4.0, 2.0] | ['EA', 'IE', 'LogVis', 'MolWt', 'n_F', 'n_O'] : [0.0, 7.0, -0.1, 250, 4.0, 2.0] |
| COc1cc(C(F)(F)F)nc(OC)c1CCF<br>CC(O)c1c(OC(F)(F)F)cc(F)cc1CN<br>OCCC(=O)Nc1ccc(F)c(C(F)(F)F)c1<br>CC(N)(C(=O)O)c1nc(C(F)(F)F)ccc1F<br>O=C(O)c1ccn(CCC(F)(F)C(F)F)c1F | OCC(O)Cc1ncc(C(F)(F)F)cc1CF<br>O=C(O)CC(CC(F)(F)C(F)F)c1ccc[nH]1<br>CCC(NCC(F)(F)C(F)F)C(=O)OCC<br>COc1ccc2c(F)c(F)c(F)c(F)c2c1O<br>NCc1cc(OC(F)(F)F)c(O)c(CF)c1C |
| ['EA', 'IE', 'LogVis', 'MolWt', 'n_F', 'n_O'] : [0.5, 7.0, -0.1, 250, 6.0, 1.0] | ['EA', 'IE', 'LogVis', 'MolWt', 'n_F', 'n_O'] : [0.0, 7.0, -0.1, 250, 6.0, 1.0] |
| OCc1nc(C(F)(F)F)c(C(F)(F)F)s1<br>OCCc1c(F)c(F)c(C(F)(F)F)c(F)c1<br>C[C@@H](O)c1c(F)c(F)c(C(F)(F)F)c(F)c1<br>OC(c1cc(F)cc(F)c1)C(F)(F)C(F)F<br>CCc1c(OC(F)(F)F)n[nH]c1C(F)(F)F | N[C@@H](CO)c1c(F)c(F)c(F)c(F)c1CF<br>CCc1ccc(OC(F)(F)F)c(C(F)(F)F)c1<br>Nc1cnc(OC(F)(F)F)c(C(F)(F)F)c1<br>OC[C@@H](c1cc(F)c(F)c(F)c1)C(F)(F)F<br>COc1ncc(C(F)(F)F)c(C(F)(F)F)n1 |
| ['EA', 'IE', 'LogVis', 'MolWt', 'n_F', 'n_O'] : [0.5, 7.0, -0.1, 250, 6.0, 2.0] | ['EA', 'IE', 'LogVis', 'MolWt', 'n_F', 'n_O'] : [0.0, 7.0, -0.1, 250, 6.0, 2.0] |
| C=C(O)CC(=O)C(C(F)(F)F)C(F)(F)CF<br>OC(F)(F)C(F)(F)Oc1ccc(F)c(F)c1<br>OC[C@@H](O)CCC(F)(F)C(F)(F)C(F)(F)F<br>OCCOCCC(F)(F)C(F)(F)C(F)(F)F<br>OCc1c(O)c(F)c(F)c(F)c1C(F)(F)F | C[Si](C)(O)OC(F)(F)C(F)(F)C(F)(F)F<br>COCC(O)CN(C(F)(F)F)C(F)(F)F<br>CCOC(C)C(O)(C(F)(F)F)C(F)(F)CF<br>O=C(O)CCCC(F)(F)C(F)(F)CC(F)(F)F<br>*--invalid--* |

| ['EA', 'IE', 'LogVis', 'MolWt', 'n_F', 'n_O'] : [0.5, 7.0, -0.1, 300, 4.0, 1.0] | ['EA', 'IE', 'LogVis', 'MolWt', 'n_F', 'n_O'] : [0.0, 7.0, -0.1, 300, 4.0, 1.0] |
|---|---|
| O=C(Cn1nc(C(F)(F)F)cc1Cl)c1ccccc1F<br>CC1CCN(C(=O)Nc2cc(F)cc(C(F)(F)F)c2)CC1<br>OC(c1cc(F)cc(F)c1)c1ccc(C(F)(F)Cl)cc1<br>OC(c1cc(F)cc(F)c1)c1cnc(C(F)(F)Cl)cc1<br>CCc1nc(-c2ccc(OC(F)(F)F)cc2)nc(C)c1F | Cc1ccc(CC(=O)Nc2cc(F)cc(F)c2)c(F)c1F<br>Cc1cc(C(F)(F)F)nc(Oc2cc(F)cc(Cl)c2)n1<br>CCCc1ncc(C(F)(F)F)c(Oc2ccc(F)cc2)n1<br>CC(OCC(F)(F)C(F)F)c1ccc(Cl)cc1Cl<br>N#Cc1ccc(OCC(F)(F)C(F)F)cc1Br |
| ['EA', 'IE', 'LogVis', 'MolWt', 'n_F', 'n_O'] : [0.5, 7.0, -0.1, 300, 4.0, 2.0] | ['EA', 'IE', 'LogVis', 'MolWt', 'n_F', 'n_O'] : [0.0, 7.0, -0.1, 300, 4.0, 2.0] |
| Cc1cc(OC(F)(F)C(F)F)C(=O)NC2CC2)cs1<br>COC(=O)c1ncc(C(F)(F)F)c(F)c1Br<br>COC(=O)Cc1nc(C(F)(F)F)c(F)cc1CCl<br>CCOC(=O)Cc1cc(C(F)(F)F)cc(F)c1CCl<br>FC(F)(F)Oc1cc(OC(F)F)cc(Br)n1 | NC(=O)COc1ccc(C(F)(F)F)c(F)c1Br<br>COC(=O)c1ccc(C(F)(F)F)c(-2ccc(F)cc2)c1<br>FC(F)(F)Oc1ccc(OCc2ccncc2)c(F)c1C<br>CCOc1cc(OC(F)(F)F)c(F)cc1Br<br>N[C@@H](CC(=O)O)c1c(C(F)(F)F)cc(F)cc1CCl |
| ['EA', 'IE', 'LogVis', 'MolWt', 'n_F', 'n_O'] : [0.5, 7.0, -0.1, 300, 6.0, 1.0] | ['EA', 'IE', 'LogVis', 'MolWt', 'n_F', 'n_O'] : [0.0, 7.0, -0.1, 300, 6.0, 1.0] |
| FC(F)C(F)(F)Oc1cc(C(F)(F)F)cnc1CCl<br>OCc1c(C(F)(F)F)ccc(C(F)(F)F)c1CCl<br>CCc1nc(OC(F)(F)F)c(C(F)(F)F)cc1CC#N<br>OC(c1nc2ccccc2s1)C(F)(F)C(F)(F)CF<br>Fc1ccc(-c2ccc(OC(F)(F)F)cc2)c(F)c1F | Nc1ccc(OCC(F)(F)C(F)(F)C(F)F)cc1C#N<br>Oc1cc(F)c(-c2ccc(C(F)(F)F)cc2)c(F)c1F<br>CCN(CC(F)C(F)(F)C(F)(F)F)C(=O)NC1CC1<br>Fc1ccc(C(F)(F)F)c(Oc2cccc(F)c2)c1F<br>Fc1cc(OC(F)(F)F)ccc1-c1ccc(F)c(F)c1 |
| ['EA', 'IE', 'LogVis', 'MolWt', 'n_F', 'n_O'] : [0.5, 7.0, -0.1, 300, 6.0, 2.0] | ['EA', 'IE', 'LogVis', 'MolWt', 'n_F', 'n_O'] : [0.0, 7.0, -0.1, 300, 6.0, 2.0] |
| Cc1c(OC(F)(F)F)cnc(C(F)(F)F)c1CC(N)=O<br>O=C(O)C(CC(F)(F)C(F)C(F)F)c1ccccc1<br>OCc1cc(OC(F)(F)F)c(Cl)cc1C(F)(F)F<br>C[C@@](N)(C(=O)O)c1nc(C(F)(F)F)c(C(F)(F)F)n1C<br>CCOC(=O)c1cc(C(F)(F)F)ccc1C(F)(F)CF | Oc1ccc(COCC(F)(F)C(F)(F)C(F)F)cc1F<br>OC(O)(Cc1ccc(C(F)(F)F)cc1)CC(F)(F)CF<br>OC(O)(c1cc(C(F)(F)F)cc(C(F)(F)F)c1)C1CC1<br>O=C(NCC(F)(F)C(F)(F)C(F)F)c1ccc(O)cc1<br>FC(F)(F)COCCOc1c(F)cc(C(F)F)cc1N |
| ['EA', 'IE', 'LogVis', 'MolWt', 'n_F', 'n_O'] : [0.5, 7.5, 0.0, 250, 4.0, 1.0] | ['EA', 'IE', 'LogVis', 'MolWt', 'n_F', 'n_O'] : [0.0, 7.5, 0.0, 250, 4.0, 1.0] |
| Cc1ccc(CNC(=O)C(F)(F)C(F)F)cc1<br>Cc1cc(CC(=O)NCC(F)(F)F)ccc1F<br>O=C(c1cccnc1)c1cc(F)c(F)c(F)c1F<br>FC(F)(F)c1cccc(Oc2ccccc2)c1F<br>CCNC(=O)Nc1ccc(F)c(C(F)(F)F)c1 | Cc1c(OC(F)(F)F)cc(F)cc1CCl<br>OC(CC(F)(F)F)c1ccc(F)c(Cl)c1<br>COc1cc(C(F)(F)F)c(F)cc1CCl<br>OCc1cnc(C(F)F)c(Cl)c1C(F)(F)F<br>Nc1cc(OCC(F)(F)C(F)F)ccc1C#N |
| ['EA', 'IE', 'LogVis', 'MolWt', 'n_F', 'n_O'] : [0.5, 7.5, 0.0, 250, 4.0, 2.0] | ['EA', 'IE', 'LogVis', 'MolWt', 'n_F', 'n_O'] : [0.0, 7.5, 0.0, 250, 4.0, 2.0] |
| OCCC(=O)Nc1c(F)cccc1C(F)(F)F<br>O=C(Cc1ccc(OC(F)(F)F)cc1)C1CC1<br>CCOc1c(OC(F)(F)F)cc(F)cc1C#N<br>CCOC1(C(F)(F)F)Oc2ccc(F)cc2C1<br>CC(C)CC(=O)NCC(O)CC(F)(F)C(F)F | OCc1cc(OCCC(F)(F)F)cc(F)c1C<br>OCc1c(C(F)F)ncc(OC(F)F)c1CC<br>COc1nc(OC(F)(F)F)c(F)cc1CC#N<br>Cc1ccc(C(F)(F)C(F)(F)C(=O)O)cc1C<br>Cc1nc(C(F)(F)F)c(CC(=O)O)cc1CF |

| ['EA', 'IE', 'LogVis', 'MolWt', 'n_F', 'n_O'] : [0.5, 7.5, 0.0, 250, 6.0, 1.0] | ['EA', 'IE', 'LogVis', 'MolWt', 'n_F', 'n_O'] : [0.0, 7.5, 0.0, 250, 6.0, 1.0] |
|---|---|
| O=C(NCC(F)(F)C(F)CF)CC(F)(F)F<br>Cc1cnc(OC(F)(F)F)c(C(F)(F)F)c1<br>O=C(Nc1cc(F)cc(F)c1F)C(F)(F)F<br>NCC(O)CCC(C(F)(F)F)C(F)(F)F<br>Fc1cc(OCC(F)(F)F)ccc1C(F)F | OC(c1cc(F)c(F)c(F)c1)CC(F)(F)F<br>FCOc1cc(C(F)(F)F)ccc1C(F)(F)F<br>Cc1ncc(C(F)(F)F)c(OC(F)(F)F)n1<br>OC(c1c(F)c(F)c(F)c1F)C(F)F<br>OCCc1c(F)c(F)c(C(F)(F)F)c(F)c1 |
| ['EA', 'IE', 'LogVis', 'MolWt', 'n_F', 'n_O'] : [0.5, 7.5, 0.0, 250, 6.0, 2.0] | ['EA', 'IE', 'LogVis', 'MolWt', 'n_F', 'n_O'] : [0.0, 7.5, 0.0, 250, 6.0, 2.0] |
| O=C(O)CCCCC(F)(F)C(F)(F)C(F)F<br>CCOC(=O)CC(C(F)(F)F)C(F)(F)CF<br>Fc1cccc(OC(F)(F)OC(F)(F)F)c1<br>Oc1cc(OC(F)(F)F)cc(C(F)(F)F)c1<br>CCC(=O)OCC(C(F)(F)F)C(F)(F)CF | FC(F)(F)C1(C(F)(F)F)OCCCCOC1<br>COC(=O)CCC(C(F)(F)F)C(F)(F)CF<br>Cc1c(O)cc(C(F)(F)F)cc1OC(F)(F)F<br>O=C(OCCCCC(F)(F)F)CC(F)(F)F<br>CC(O)CCC(O)(C(F)(F)F)C(F)(F)F |
| ['EA', 'IE', 'LogVis', 'MolWt', 'n_F', 'n_O'] : [0.5, 7.5, 0.0, 300, 4.0, 1.0] | ['EA', 'IE', 'LogVis', 'MolWt', 'n_F', 'n_O'] : [0.0, 7.5, 0.0, 300, 4.0, 1.0] |
| O=C(NCC(F)(F)C(F)F)c1ccc(Cl)cc1Cl<br>CCc1ncc(OC(F)(F)F)c(F)c1CBr<br>FC(F)(F)c1ccc(Oc2ncccc2Cl)c(F)c1<br>Nc1ncc(F)cc1C(=O)Nc1ccc(C(F)(F)F)cc1<br>CCCc1nc(OC(F)(F)F)nc(F)c1Br | CN(C)c1cc(C(F)(F)F)nc(Oc2ccc(F)cc2)n1<br>OC(c1cccc(C(F)(F)F)c1)c1c(F)cccc1Cl<br>Cc1cnc(C(F)(F)F)c(Oc2ccc(F)c(Cl)c2)n1<br>N#Cc1ccc(COc2ccccc2C(F)(F)F)c(F)c1<br>CNc1ccc(Oc2ccc(C(F)(F)F)nc2)c(F)c1C |
| ['EA', 'IE', 'LogVis', 'MolWt', 'n_F', 'n_O'] : [0.5, 7.5, 0.0, 300, 4.0, 2.0] | ['EA', 'IE', 'LogVis', 'MolWt', 'n_F', 'n_O'] : [0.0, 7.5, 0.0, 300, 4.0, 2.0] |
| CC(Oc1cccc(C(F)(F)F)c1)C(=O)c1ccccc1<br>COCc1nc(OC(F)(F)F)c(F)cc1Br<br>O=C(NCC(F)(F)CO)c1c(F)cc(F)cc1CCl<br>FC(F)(F)Oc1cc(F)c(OCc2cccnc2)c(C)c1<br>CS(=O)(=O)Nc1ccc(C(F)(F)C(F)F)c(Cl)c1 | CS(=O)(=O)N1CCN(c2c(F)c(F)cc(F)c2F)C1<br>Oc1cc(OC(F)(F)F)ccc1CBr<br>CS(=O)(=O)Nc1ccc(SC(F)(F)C(F)F)cc1<br>O=S(=O)(c1ccccc(C(F)(F)F)c1)c1ccc(F)cc1<br>CC(Nc1cc(C(F)(F)F)cc(F)c1)C(=O)OC(C)(C)C |
| ['EA', 'IE', 'LogVis', 'MolWt', 'n_F', 'n_O'] : [0.5, 7.5, 0.0, 300, 6.0, 1.0] | ['EA', 'IE', 'LogVis', 'MolWt', 'n_F', 'n_O'] : [0.0, 7.5, 0.0, 300, 6.0, 1.0] |
| Cc1c(C(F)(F)F)cc(OC(F)(F)F)nc1CCl<br>Nc1cc(OC(F)(F)F)cc(C(F)(F)F)c1CCl<br>FC(F)(F)Oc1ccc(-c2ccc(C(F)(F)F)cc2)cc1<br>O=Cc1ccc(C(F)(F)F)c(C(F)(F)F)c1CCl<br>--invalid-- | Oc1c(F)cccc1-c1cc(C(F)(F)F)cc(F)c1F<br>OC(c1cccc(C(F)(F)F)c1)c1cccc(F)c1F<br>CC(=O)Nc1c(C(F)(F)F)cnc(C(F)(F)F)c1CN<br>OCc1cc(C(F)(F)F)cc(C(F)(F)F)c1CCl<br>OC(c1ccc(C(F)(F)F)nc1)c1c(F)cc(F)cc1F |
| ['EA', 'IE', 'LogVis', 'MolWt', 'n_F', 'n_O'] : [0.5, 7.5, 0.0, 300, 6.0, 2.0] | ['EA', 'IE', 'LogVis', 'MolWt', 'n_F', 'n_O'] : [0.0, 7.5, 0.0, 300, 6.0, 2.0] |
| OCC(Oc1cccc(C(F)(F)F)c1)CC(F)(F)CF<br>O=Cc1cc(OCC(F)(F)C(F)(F)F)ccc1N<br>OCc1c(OC(F)(F)F)ncc(C(F)(F)F)c1Cl<br>NC(COCC(F)(F)F)c1cccc(OC(F)(F)F)c1<br>OCc1ccccc1OCCC(F)(F)C(F)(F)C(F)F | C[C@@H](NC(=O)O)c1cc(C(F)(F)F)cc(C(F)(F)F)c1<br>CCOC(=O)Cc1c(F)cc(C(F)(F)F)cc1C(F)F<br>NC(=O)c1c(OC(F)(F)F)cnc(C(F)(F)F)c1CN<br>O=C(NCC(F)(F)F)N1CCC(O)(C(F)(F)F)CC1<br>FC(F)(F)Oc1ccc(OC(F)(F)C(F)Cl)cc1 |

| ['EA', 'IE', 'LogVis', 'MolWt', 'n_F', 'n_O'] : [0.5, 7.5, -0.1, 250, 4.0, 1.0] | ['EA', 'IE', 'LogVis', 'MolWt', 'n_F', 'n_O'] : [0.0, 7.5, -0.1, 250, 4.0, 1.0] |
|---|---|
| CNCC(=O)Nc1cc(F)c(C(F)(F)F)cc1<br>OC(CC(F)(F)F)c1ccc(F)c(Cl)c1<br>Fc1ccc(COc2ccccc2F)c(F)c1F<br>OCc1cc(C(F)(F)F)c(F)cc1CCl<br>COc1c(C(F)(F)F)ccc(F)c1CCl | OC(c1cc(F)cc(C(F)(F)F)c1)C1CCC1<br>OCCc1cc(C(F)(F)F)c(Cl)cc1F<br>OCCc1nc(C(F)(F)F)c(F)cc1Cl<br>Nc1cnc(OCCCC(F)(F)F)c(F)c1F<br>Nc1ccc(OCCC(F)C(F)F)c(C)c1 |
| ['EA', 'IE', 'LogVis', 'MolWt', 'n_F', 'n_O'] : [0.5, 7.5, -0.1, 250, 4.0, 2.0] | ['EA', 'IE', 'LogVis', 'MolWt', 'n_F', 'n_O'] : [0.0, 7.5, -0.1, 250, 4.0, 2.0] |
| CN(C(=O)O)c1cc(C(F)(F)F)cc(F)c1C<br>CS(=O)(=O)CCSCC(F)(F)C(F)F<br>CCC(=O)NCC(=O)NCC(F)(F)C(F)F<br>CNC1CC(C(F)(F)C(F)(F)C(=O)O)CC1<br>COC(=O)Nc1cccc(C(F)(F)C(F)F)c1 | CCCC(=O)OCCCCC(F)(F)C(F)(F)F<br>O=Cc1cc(OCC(F)C(F)F)cs1<br>OC(COCC(F)(F)F)c1ccc(F)c(C)c1<br>CCCOC(=O)Nc1c(F)cc(F)c1F<br>CCOC(=O)Nc1nc(C(F)(F)F)ccc1F |
| ['EA', 'IE', 'LogVis', 'MolWt', 'n_F', 'n_O'] : [0.5, 7.5, -0.1, 250, 6.0, 1.0] | ['EA', 'IE', 'LogVis', 'MolWt', 'n_F', 'n_O'] : [0.0, 7.5, -0.1, 250, 6.0, 1.0] |
| C[C@H](O)c1cc(C(F)(F)F)cc(C(F)(F)F)c1<br>OCC(F)C(F)(F)c1cc(F)cc(F)c1<br>C[C@H](N)CC(=O)NC(C(F)(F)F)C(F)(F)F<br>Cc1cc(OC(F)(F)F)nc(C(F)(F)F)c1<br>--invalid-- | Cc1c(OC(F)(F)F)cccc1C(F)(F)F<br>Nc1ncc(F)c(OC(F)(F)F)c1C(F)F<br>CNCC(=O)NCC(F)(F)C(F)(F)C(F)F<br>Fc1cc(F)c(OCCC(F)(F)F)c(F)c1<br>FC(F)(F)CCOc1cc(F)cc(F)c1 |
| ['EA', 'IE', 'LogVis', 'MolWt', 'n_F', 'n_O'] : [0.5, 7.5, -0.1, 250, 6.0, 2.0] | ['EA', 'IE', 'LogVis', 'MolWt', 'n_F', 'n_O'] : [0.0, 7.5, -0.1, 250, 6.0, 2.0] |
| OCc1c(F)c(F)c(OC(F)(F)F)c(F)c1<br>C=CCOC(=O)CC(F)(F)C(F)(F)C(F)F<br>OCc1cc(C(F)(F)F)cc(C(F)(F)F)c1O<br>Cc1c(OC(F)(F)F)[nH]c(C(F)(F)F)c1O<br>C=CCOC(=O)C(C(F)(F)F)C(F)(F)CF | COC(=O)CCC(C(F)(F)F)C(F)(F)CF<br>CC(CC(F)(F)C(F)(F)C(F)F)CC(=O)O<br>CC(C)(CC(F)(F)C(F)(F)C(F)F)C(=O)O<br>Cc1ccc(OC(F)(F)F)c(OC(F)(F)F)c1<br>FCOc1c(C(F)(F)F)[nH]c(C(F)F)c1O |
| ['EA', 'IE', 'LogVis', 'MolWt', 'n_F', 'n_O'] : [0.5, 7.5, -0.1, 300, 4.0, 1.0] | ['EA', 'IE', 'LogVis', 'MolWt', 'n_F', 'n_O'] : [0.0, 7.5, -0.1, 300, 4.0, 1.0] |
| OC(c1c(F)c(F)c(F)c(F)c1Br)C1CC1<br>Oc1nc(F)c(C(F)(F)F)cc1I<br>OC(Cc1ccc(F)c(Br)c1)CC(F)(F)F<br>COc1nc(C(F)(F)F)ccc1CBr<br>O=C(Cc1nc(C(F)(F)F)ns1)c1ccc(F)cc1C | OC(c1cccc(C(F)(F)F)c1)c1ccc(F)cc1Cl<br>CC1CC(c2ccc(C(F)(F)F)cc2)NC(=O)NC1F<br>CC(NC(=O)CC(F)(F)C(F)(F)F)c1ccc(Cl)cc1<br>COc1cc(Br)cc(CC(F)(F)C(F)(F)F)c1<br>Cn1nc(C(F)(F)F)cc1C(=O)Nc1ccc(F)cc1C |
| ['EA', 'IE', 'LogVis', 'MolWt', 'n_F', 'n_O'] : [0.5, 7.5, -0.1, 300, 4.0, 2.0] | ['EA', 'IE', 'LogVis', 'MolWt', 'n_F', 'n_O'] : [0.0, 7.5, -0.1, 300, 4.0, 2.0] |
| CC(=O)Nc1nc(-c2ccc(C(F)(F)F)c(F)c2)c(C)o1<br>Oc1cccc(Oc2c(F)c(F)c(F)c(F)c2Cl)c1C<br>O=S(=O)(Cc1ccc(F)cc1)c1c(F)cc(F)cc1F<br>OC(O)(c1c(F)cccc1F)c1cc(F)c(Cl)cc1F<br>--invalid-- | CC(Oc1cccc(C(F)(F)F)c1)c1cccc(F)c1O<br>COc1nc(C(F)(F)F)ccc1OCc1ccc(F)cc1<br>OC(COCC(F)(F)F)Cc1ncccc1Cl<br>FC(F)C(F)(F)Oc1cccc1COc1cccc1<br>COC(=O)c1ncc(C(F)(F)F)cc1-c1ccc(F)cc1 |

| ['EA', 'IE', 'LogVis', 'MolWt', 'n_F', 'n_O'] : [0.5, 7.5, -0.1, 300, 6.0, 1.0] | ['EA', 'IE', 'LogVis', 'MolWt', 'n_F', 'n_O'] : [0.0, 7.5, -0.1, 300, 6.0, 1.0] |
|---|---|
| OC(C(F)(F)C(F)(F)C(F)(F)F)C(Cl)(Cl)Cl<br>OCC(c1c(F)cc(F)cc1Cl)C(F)(F)C(F)(F)F<br>FC(F)(F)Oc1ccnc(-c2ccc(C(F)(F)F)cc2)c1<br>CS(=O)c1cc(C(F)(F)F)cc(C(F)(F)F)c1C#N<br>OC(F)(C(F)(F)F)C(F)(F)I | OCC(c1ccc(C(F)(F)F)c(C(F)(F)F)c1)C1CC1<br>FC(F)(F)Oc1cc(C(F)(F)F)c(Cl)cc1Cl<br>Cc1ccc(OCC(F)(F)C(F)(F)C(F)(F)Cl)cc1<br>Fc1cc(OC(F)(F)F)ccc1-c1cc(F)cc(F)c1<br>--invalid-- |
| ['EA', 'IE', 'LogVis', 'MolWt', 'n_F', 'n_O'] : [0.5, 7.5, -0.1, 300, 6.0, 2.0] | ['EA', 'IE', 'LogVis', 'MolWt', 'n_F', 'n_O'] : [0.0, 7.5, -0.1, 300, 6.0, 2.0] |
| FC(F)(F)c1cc2c(cc1C(F)(F)F)OCCCCO2<br>CCCC(=O)OCCCCC(F)(F)C(F)(F)C(F)(F)F<br>COC(=O)Cc1cc(C(F)(F)F)cc(C(F)(F)F)c1C<br>O=C(Cc1cc(C(F)(F)F)cc(C(F)(F)F)c1)C(N)=O<br>CCc1cc(C(F)(F)F)cc(C(F)(F)F)c1CC(=O)O | OC(Cc1ccc(OC(F)(F)F)cc1)CC(F)(F)CF<br>OCc1ccc(COCC(F)(F)C(F)(F)C(F)(F)F)cc1<br>Oc1cc(OC(F)(F)F)ccc1SCC(F)(F)F<br>NC(=O)Cc1cc(C(F)(F)F)nc(OC(F)(F)F)c1C<br>OC(c1c(F)c(F)c(F)c1F)c1ccc(F)cc1O |

# Conclusion

In summary, we demonstrate a novel conditional molecule generation algorithm ConGen, which is based on semi-supervised variational auto-encoder (SSVAE) technology. However, unlike the SSVAE model, the ConGen model is explicitly designed such that it can work with dirty training data with incomplete labels. This is important because in practice, the molecules we can find from publicly available databases or characterized by in-house simulations and experiments will have incomplete sets of properties available, due to various factors (intellectual property, commercial secret, etc) as well as experimental or computational cost considerations. A user of the baseline SSVAE model will need to remove molecules with incomplete labels or assign dummy labels on the molecules at the cost of lower model accuracy. On the other hand, the ConGen model can easily mix dirty training datasets from multiple sources. ConGen is also designed with flexibility for substituting its sub-models with other types of models, enabling the user to include pre-trained models which can be helpful, especially in cases where there is limited training data availability. Finally, we demonstrate the practical use of our model for generating the virtual screening chemical space for Li-ion battery LHCE diluent candidates with multiple co-constraint requirements.

## Experimental methods

### Ab-initio EA and IA validation

These molecule EA and IE calculations are conducted with PySCF's implementation[23,24] of the DFT Kohn-Sham method at the PBE6-31+G* level[25] with Grimme's dispersion correction[26] for geometry optimizations on the gas-phase and B3LYP/6-31+G* level of theory with the solvation energy corrections of tetrahydrofuran (THF) using the integral equation formalism polarizable continuum model (IEF-PCM)[27] implicit solvation model for single point energies. The vibrational frequencies were computed at the same level of theory at 298.15K as for the geometry optimizations to confirm whether each optimized stationary point is an energy minimum. Here, we optimize the geometry at different charge states (cation, anion, neutral) to calculate the adiabatic IE/EA.

## Data availability

All the training data sources, as well as all the structural and computational validation of the unconditionally and conditionally generated molecules are available in the ESI.

## Author contributions

J.P.M. conceptualized this project, developed the ConGen model, and performed and analysed the model training and molecule generation experiments. S.Z. supervised the project. X.L. performed ab-initio computational validation for the generated molecules. J.Q. supported model training optimization efforts. The manuscript was drafted by J.P.M., and reviewed by all the authors.

## Conflicts of interest

There are no conflicts to declare.

## Acknowledgements

The computation efforts in this work were performed in Tencent Cloud platform.